\documentclass{PoS}

\title{Measurement of neutral current $\mathbf{e^\pm p}$ cross sections at high Bjorken $\mathbf{x}$  with the ZEUS detector}

\ShortTitle{High $x$}

\author{\speaker{Aharon LEVY}
\\On behalf of the ZEUS collaboration\\
        Tel Aviv University, Tel Aviv, Israel\\
        E-mail: \email{levyaron@post.tau.ac.il}}

\abstract{The latest results of the ZEUS collaboration on the high $Q^2$, the exchanged boson virtuality, high Bjorken $x$ region up to values of $x\cong 1$ are presented. Differential cross sections in $x$ and $Q^2$  are given for $Q^2 \geq 725\ {\rm GeV}^2$. An improved reconstruction method and a substantially increased amount of data allow a finer binning in the high-$x$ region of the neutral current cross section and lead to a  measurement with much  improved precision compared to a similar earlier analysis.
The measurements are compared to Standard Model expectations based on a variety of recent parton distribution functions.  }

\FullConference{XXII. International Workshop on Deep-Inelastic Scattering and Related Subjects,\\
		28 April - 2 May 2014\\
		Warsaw, Poland}
\newcommand\units{\,\mathrm}
\usepackage{subfig}

\begin{document}

\section{Introduction}

HERA was a high-energy electron\footnote{Here and in the following the term electron denotes generically both the electron and the positron.}-proton collider, at a centre-of-mass (cms) energy of 320 GeV. It started operating in 1992 and was closed in 2007.  Due to the accessible high values of virtuality, $Q^2$,  of the exchanged boson, reaching values up to about 40 000 GeV$^2$, it could 'look' into the proton (see Fig. 1) with a resolution $\lambda$ of about 10$^{-3}$ fm.

\begin{figure}[h!]
\begin{minipage}{0.25\linewidth}
\centerline{\includegraphics[width=0.7\linewidth]{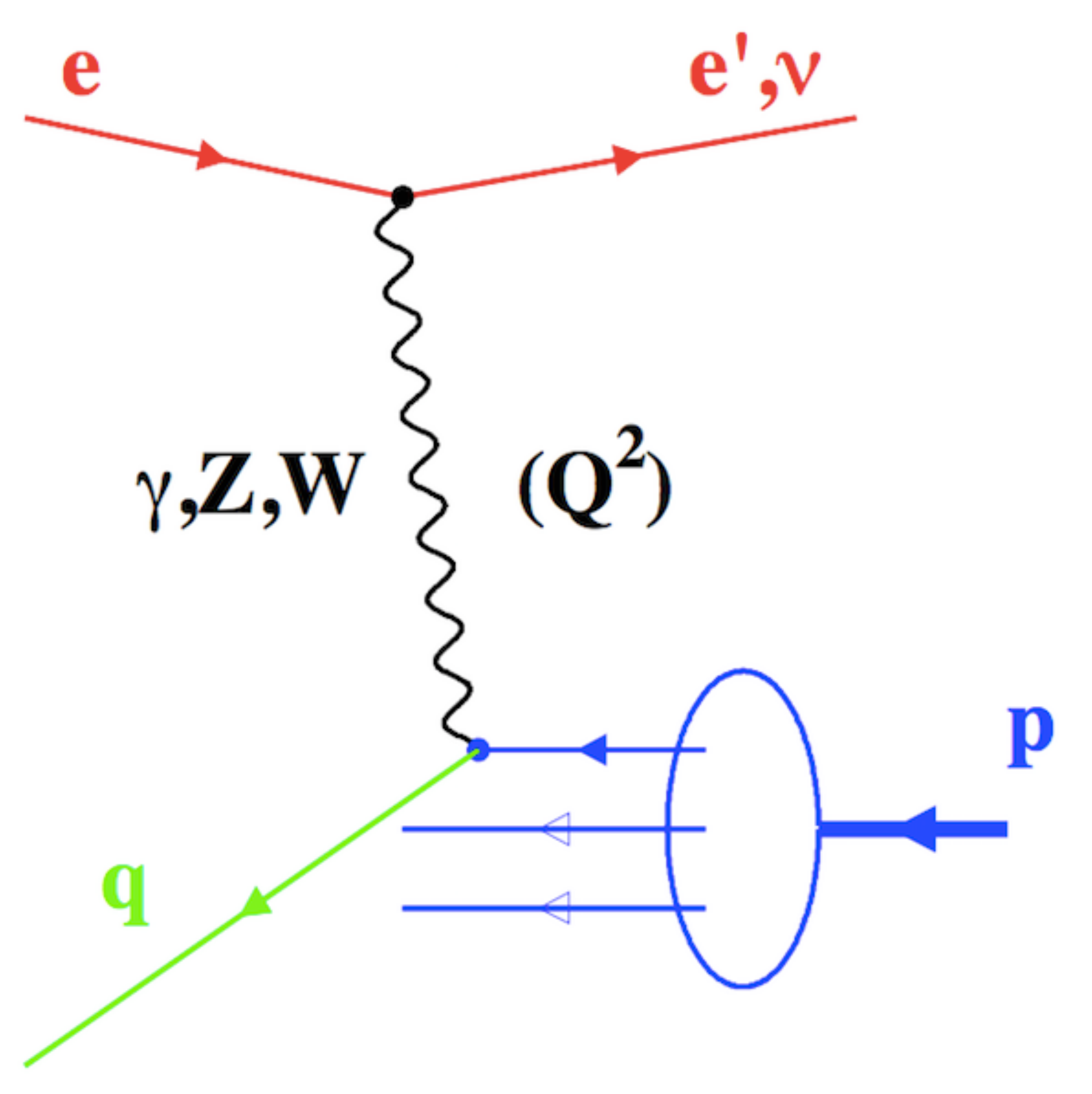}}
\vspace{-0.5cm}
\caption{Diagram describing $ep$ collisions.}
\end{minipage}
\hfill
\begin{minipage}{0.7\linewidth}
At HERA many experiments were performed by selecting events with virtualities of the exchanged photon from almost-real photons ($Q^2 \sim$ 0), the photoproduction region, through the start of the deep inelastic scattering (DIS) region, $Q^2 \sim$ 4 GeV$^2$ ($\lambda$=0.1 fm), to the very high-$Q^2$ region, $Q^2 \sim$ 40 000 GeV$^2$ ($\lambda=10^{-3}$ fm), where electroweak physics could be studied.
\end{minipage}
\end{figure}

In this talk, a recent result concerning the proton structure will be presented. It was carried out by the ZEUS collaboration~\cite{zeushighx} measuring neutral current (NC) e$^\pm$p cross sections at high $Q^2$ in the high Bjorken $x$ region up to values of $x\cong 1$.

\section{High x, extending to $x\cong 1$}

\begin{figure}[h!]
\begin{minipage}{0.38\linewidth}
\includegraphics[width=0.95\linewidth]{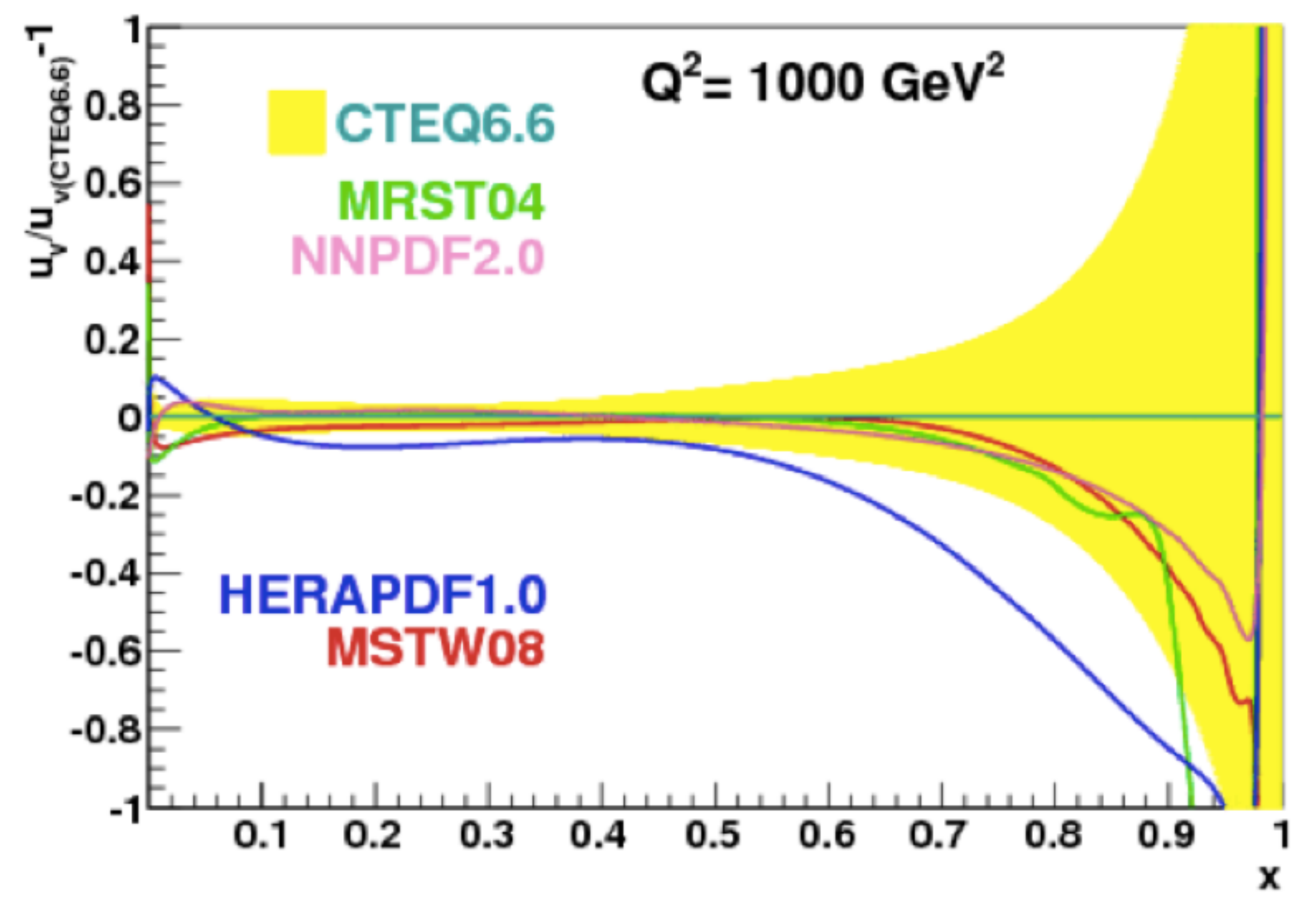}
\vspace{-0.5cm}
\caption{Example of the sizable differences between some parameterisation description of the $u$ valence quark, $u_V$.}
\label{fig:motivation}
\end{minipage}
\hfill
\begin{minipage}{0.6\linewidth}
The DIS cross sections have been measured by both collaborations with very high precision. These measurements were combined and produced text-book results with even higher precision~\cite{combined}. Nevertheless, the highest $x$ value for which measurements were done was 0.65. There are fixed-target experiments~\cite{pl:b223:485,pl:b282:475,jferson} which measure higher values of $x$ but in a low $Q^2$ region. In global perturbative quantum chromodynamic fits of parton distribution functions (PDFs), a parameterisation of the form $(1 - x)^\beta$ is assumed in order to extend PDFs ro $x$ = 1. Although all fitters use the same parameterisation, sizeable differences are obtained in the high-$x$ region~\cite{allen-eps}, as shown in Fig.~\ref{fig:motivation}.
\end{minipage}
\end{figure}

\begin{figure}[h!]
\begin{center}
\includegraphics[width=0.75\linewidth]{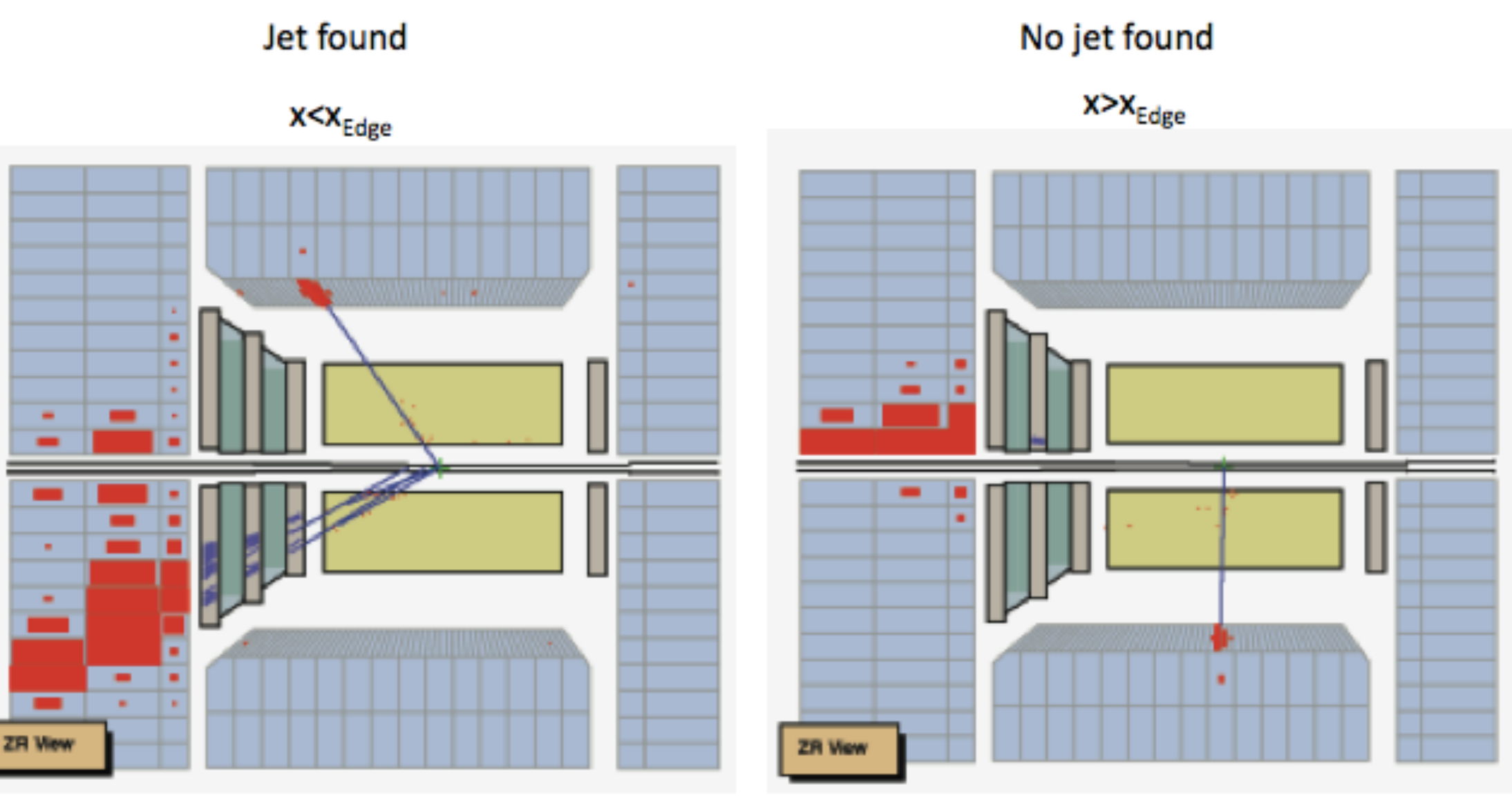}
\caption{Left-hand side: a one-jet event with a scattered electron in the BCAL and the jet fully contained in the FCAL.Also seen in FCAL are the proton remnent. Right-hand side: A zero-jet event where the scattereed electron is in BCAL and the jet remains inside the beam pipe. The proton remnant and possibly some energy emerging from the jet in the beampipe are seen in FCAL.}
\label{fig:xedge}
\end{center}
\end{figure}

The ZEUS collaboration showed in an earlier publication~\cite{epj:c49:523-544} that the kinematics of HERA and the design of the detectors allow an extension of the measurements of the neutral current (NC) cross sections up to $x$ = 1. The results presented here are based on a much larger data sample and an improved analysis procedure.

A typical NC high-$Q^{2}$ and high-$x$ event consists of the scattered electron and a high-energy collimated jet of particles in the direction of the  struck quark. The electron and the jet are balanced in transverse momentum. The proton remnant mostly disappears down the beam pipe. The $x$ and $Q^2$ of events, in which the jet is well contained in the detector, may be determined by various techniques. However,  the maximum $x$ value that can be reached is limited by the fact that at the low values of $y$ typical of these events, the uncertainty on $x=Q^2/ys$  increases as $\Delta x\sim \Delta y/y^2$. An improved $x$ reconstruction is achieved by observing that, in the limit of $x\rightarrow 1$, the energy of the struck quark represented by a collimated jet is $E_\mathrm{jet} \cong xE_p$. The  expression for $x$ is 
\begin{equation}
x = \frac {E_\mathrm{jet}(1+\cos \theta_\mathrm{jet})}{2 E_p \left( 1- \frac {E_\mathrm{jet}(1-\cos\theta_\mathrm{jet})}{2E_{e}} \right) } \, ,
\label{eq-xpt}
\end{equation}
where $\theta_\mathrm{jet}$ is the scattering angle of the jet in the detector. 

As $x$ increases and the jet associated with the struck quark disappears down the beam-pipe (see Fig.~\ref{fig:xedge}), the ability to reconstruct $x$ is limited by the energy loss. However, in these events, the cross section integrated  from a certain limit in $x$, $x_\mathrm{edge}$, up to $x=1$ is extracted. The value of $x_\mathrm{edge}$  below which the jet is fully contained in the detector depends on $Q^2$ and the higher the $Q^2$, the higher the value of $x_\mathrm{edge}$.

In the analysis presented here, two improvements were introduced~\cite{ronenphd,inderpalphd,rituphd}. Since, in these events, the scattered electron is very well measured in the detector, for the one-jet events, which statistically dominate the high-$Q^{2}$ and high-$x$ NC samples, the measured value of $E_\mathrm{jet}$ in Eq.~(\ref{eq-xpt}) is replaced by
\begin{equation}
E_{\mathrm{jet}} =  E_{T}^{{e}} / \sin \theta _{\mathrm {jet}}  \, ,
\end{equation}
as the transverse energy of the (massless) jet is balanced by the transverse energy of the scattered electron, $E_{T}^{{e}}$, and the uncertainty on the electron-energy scale and resolution is smaller than that for the hadronic component. Hard QCD processes present in DIS, such as boson-gluon fusion and QCD Compton, may lead to extra jets in the events. The analysis is thus extended to include multi-jet events. For the latter, the smallest bias on the $x$ reconstruction was found for a modified Jacquet-Blondel (JB) method~\cite{proc:epfac:1979:391}, in which 
\begin{equation}
x=\frac{(E_T^\mathrm{jets})^2}{s (1-y_\mathrm{jets})y_\mathrm{jets}} \, ,
\end{equation}
where $E_T^\mathrm{jets}$ is the vector sum over the transverse-energy vectors of individual jets and $y_\mathrm{jets}=\sum_\mathrm{jets}E_i(1-\cos \theta_i)/2E_e$, where $E_i$ and $\theta_i$ denote the energy and scattering angle of jet $i$ and the sum runs over all jets. This approach is less sensitive to the contribution of particles that are not assigned to any jets.

The value of  $Q^2$ is calculated from the measured scattered-electron energy, $E^{'}_e$, and scattering angle, $\theta_e$, as
\begin{equation}
\label{q2formula}
  Q^{2}=2E_{e} E^{'}_{e} (1+\cos\theta_e) \, .
\end{equation}
This provides the best resolution for non-radiative events.

\begin{figure}[h!]
\begin{center}
\subfloat[]{\includegraphics[height=0.11\textheight]{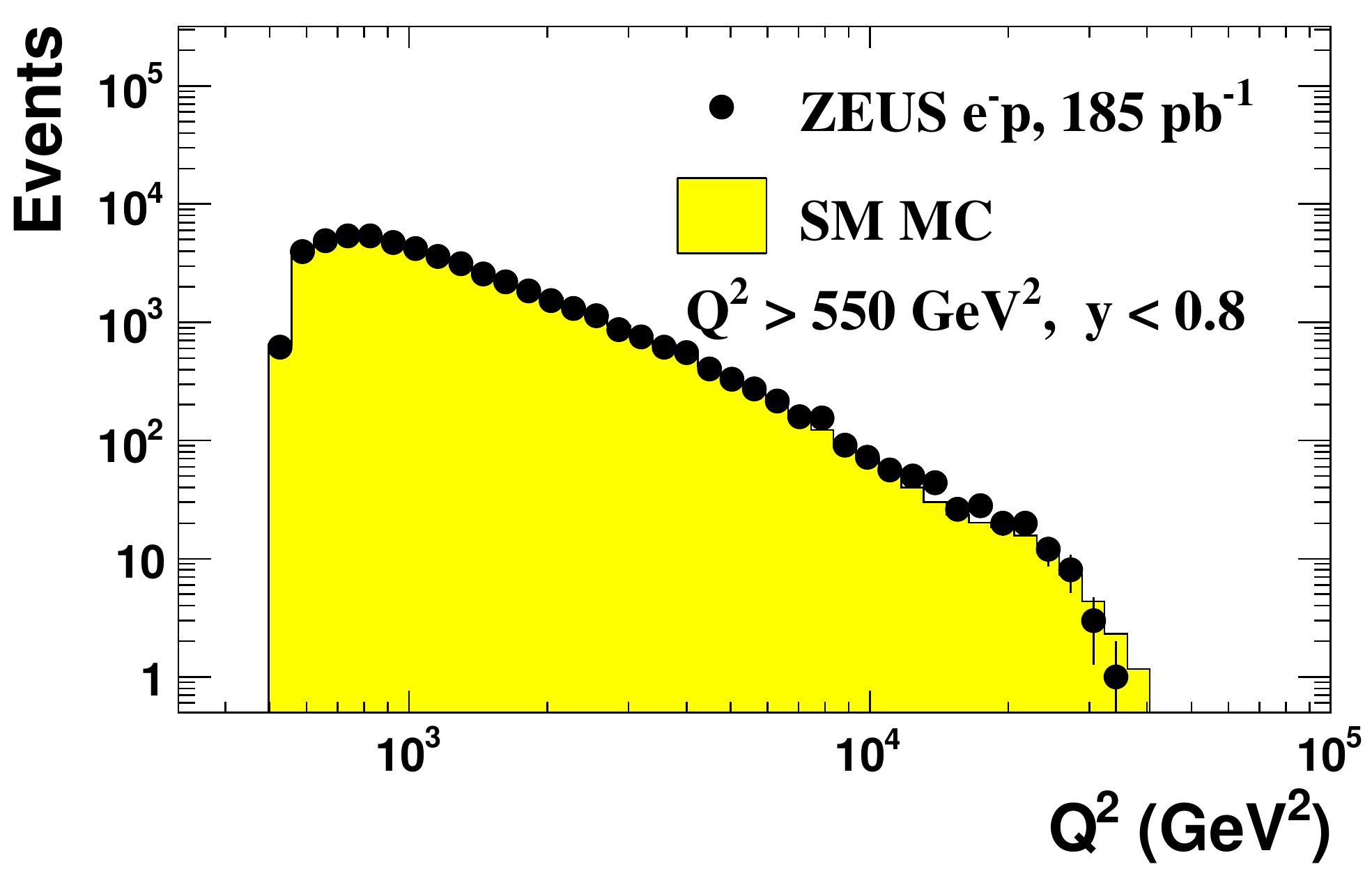}}
\subfloat[]{\includegraphics[height=0.11\textheight]{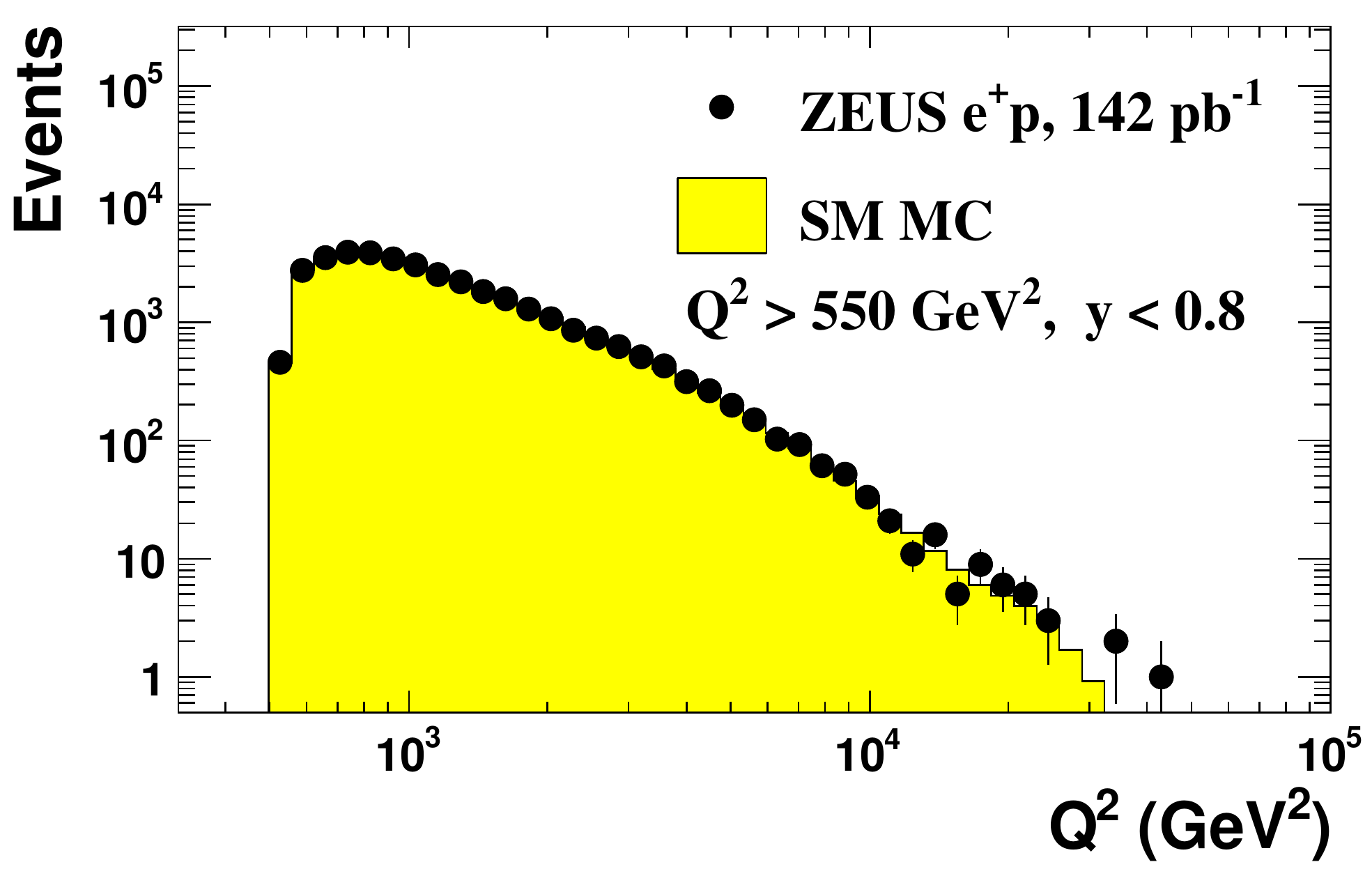}}\\
\subfloat[]{\includegraphics[height=0.11\textheight]{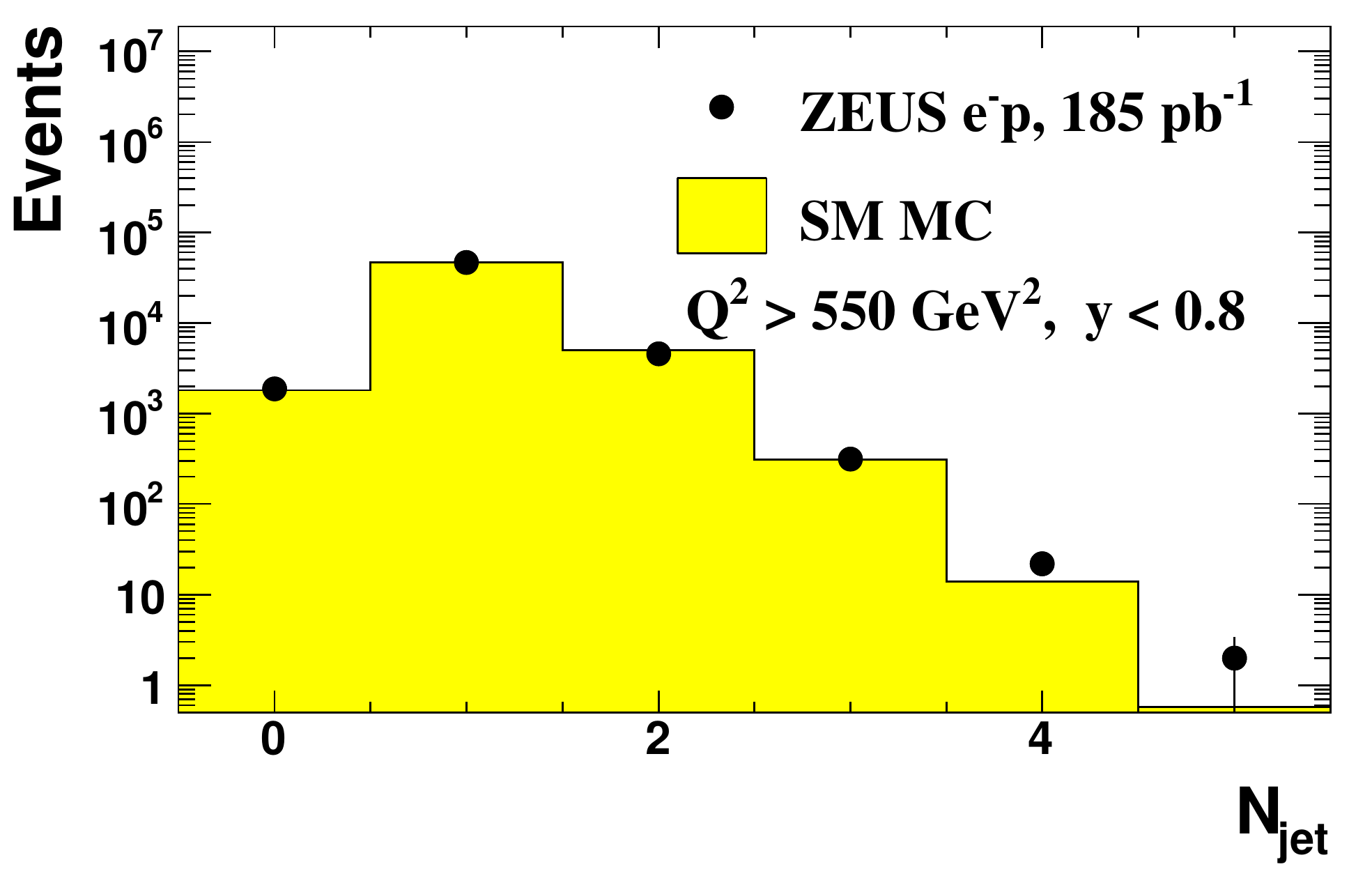}}
\subfloat[]{\includegraphics[height=0.11\textheight]{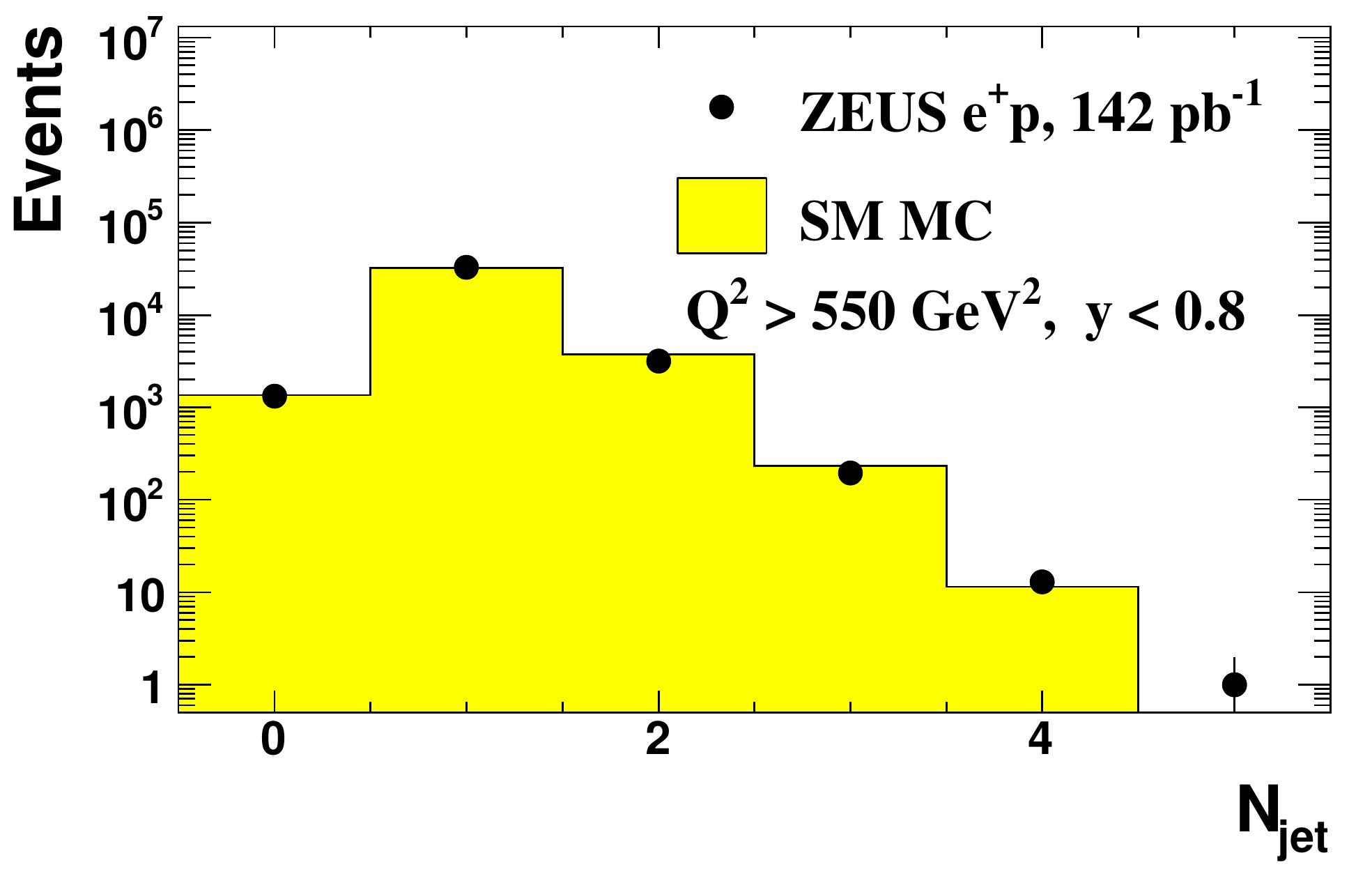}}\\
\subfloat[]{\includegraphics[height=0.11\textheight]{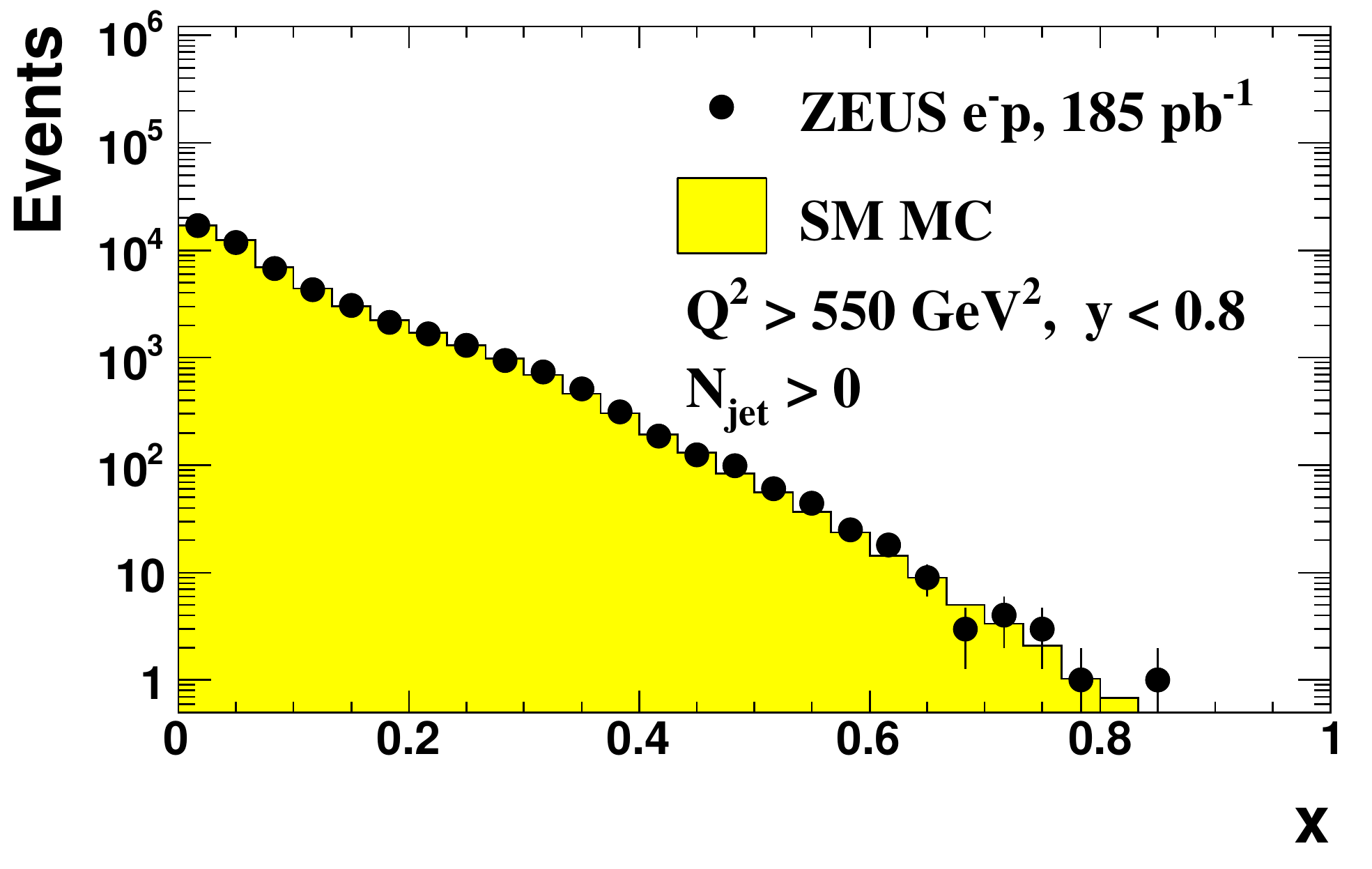}}
\subfloat[]{\includegraphics[height=0.11\textheight]{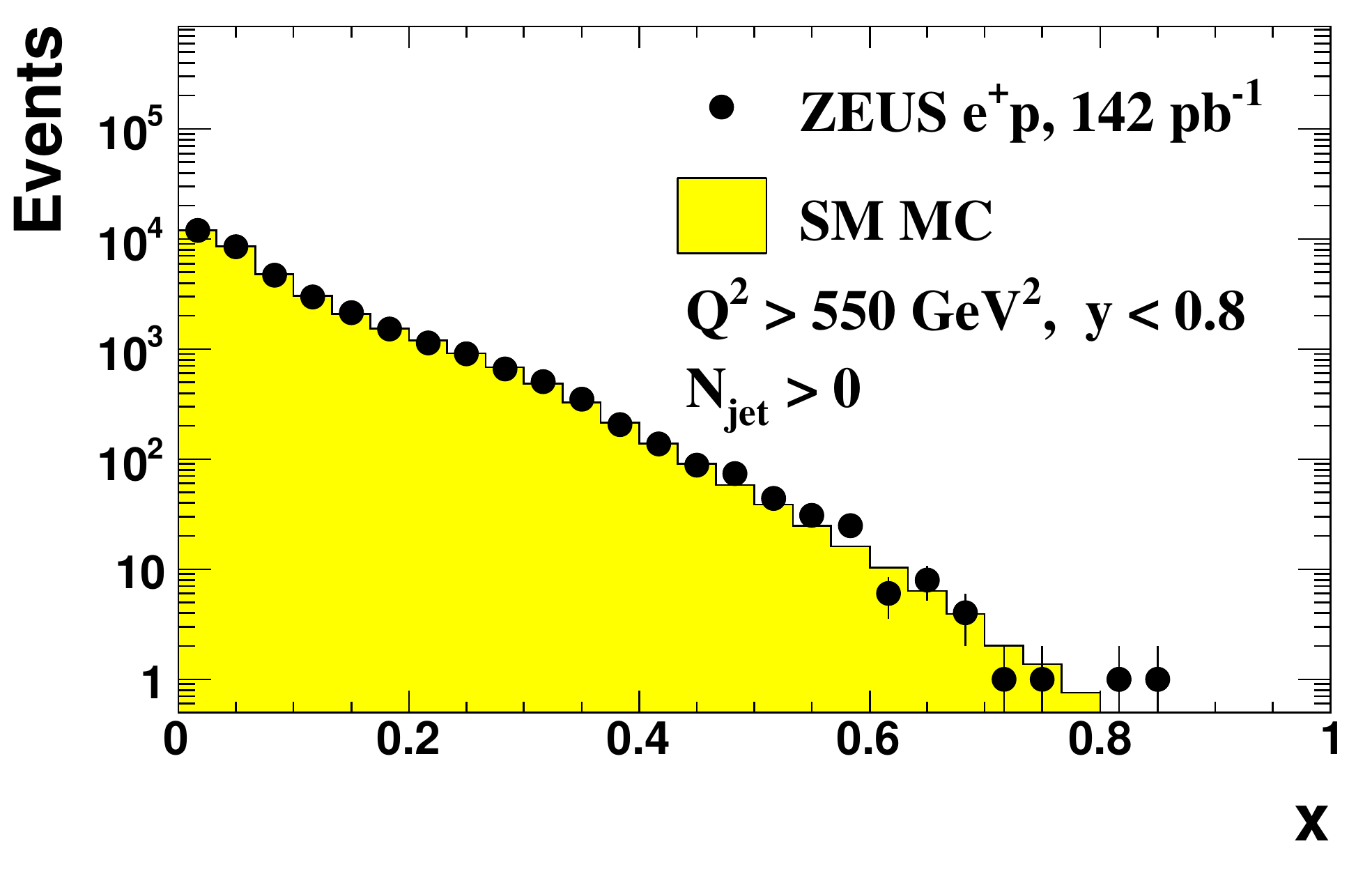}}
\caption{Distribution of  (a), (b) $ Q^2$,  (c), (d)  the jet multiplicity, $N_\mathrm{jet}$ and (e), (f)  $x$ for events with at least one jet for the $e^-p$ and $e^+p$  data samples (dots)  for $Q^2>550 \units{GeV^2}$ and $y<0.8$. The distributions are compared to MC expectations (histograms) normalised to the number of events in the data.  }
\label{fig:xq2mix}
\end{center}
\end{figure}

A comparison between data distributions and MC expectations for events with $Q^2>550 \units{GeV^2}$ and $y<0.8$, for $Q^2$,  $N_\mathrm{jet}$ and  $x$ for events with at least one reconstructed jet is shown in Fig.~\ref{fig:xq2mix}.  Very good agreement is achieved in all distributions. The MC expectations are a linear combination of {\sc Ariadne} and  {\sc Lepto}.

\section{Results}

The double-differential Born-level cross sections as a function of $Q^2$ and $x$ could be measured in many bins in $x$ because of the large number of events  in this analysis (53~099 for the $e^- p$ and 37~361 for the $e^+ p$ sample).
\begin{figure}[h!]
\begin{minipage}{0.48\linewidth}
\includegraphics[width=0.95\linewidth]{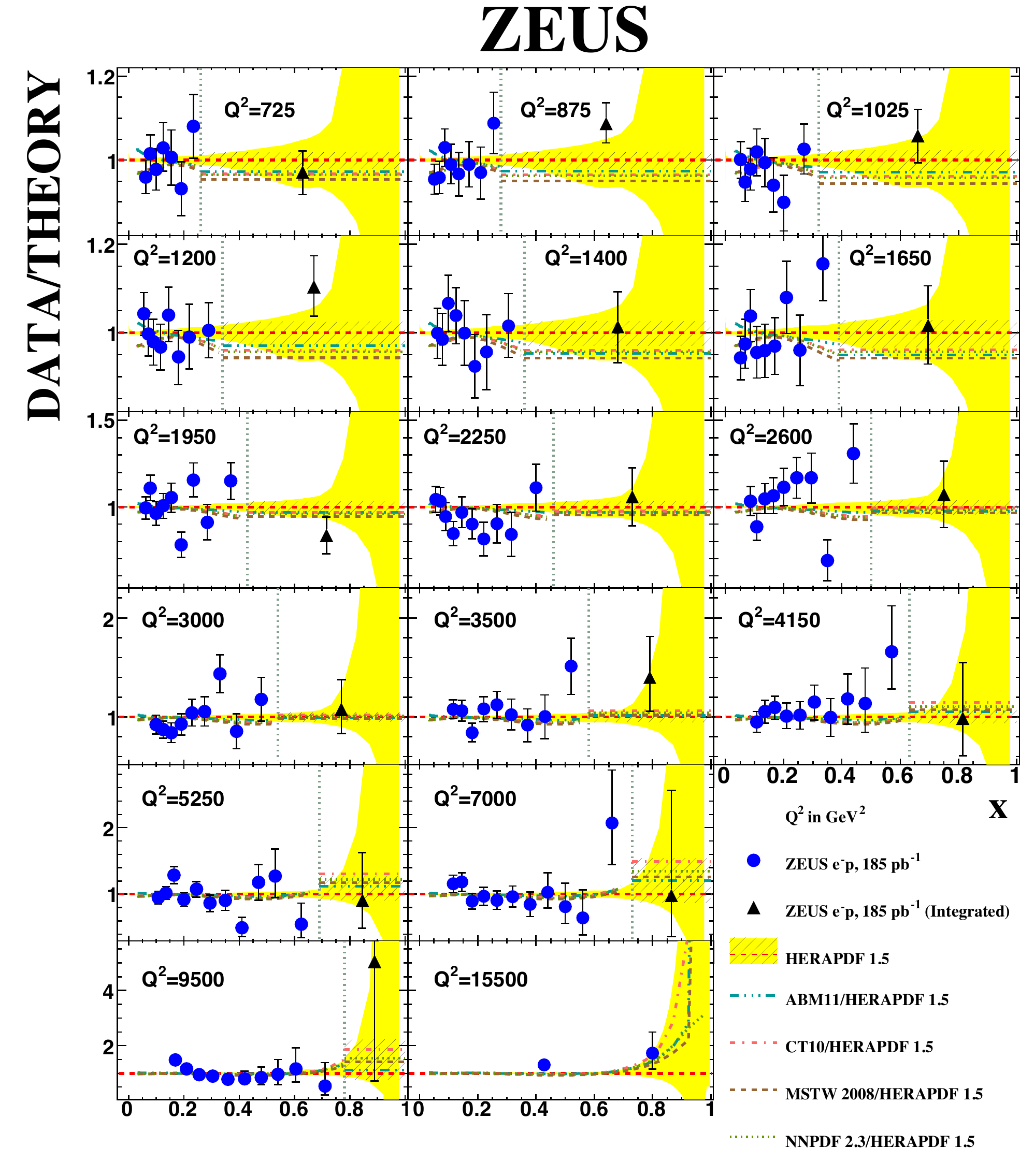}
\caption{
Ratio of the double-differential cross section for NC $e^-p$ 
scattering and  of the double-differential
cross section integrated over $x$ 
to the Standard Model expectation evaluated using the HERAPDF1.5 PDFs as a function of $x$ at different $Q^2$ values as described in the legend.  For HERAPDF1.5, the uncertainty is given as a band. The expectation for the integrated bin is also shown as a hatched box. 
The error bars show the statistical and systematic uncertainties added in quadrature.
The expectations of other commonly used PDF sets normalised to HERAPDF1.5 PDFs are also shown, as listed in the legend. Note that the scale on the $y$ axis changes with $Q^2$. }
\label{fig:eMp}
\end{minipage}
\hfill
\begin{minipage}{0.48\linewidth}
\includegraphics[width=0.95\linewidth]{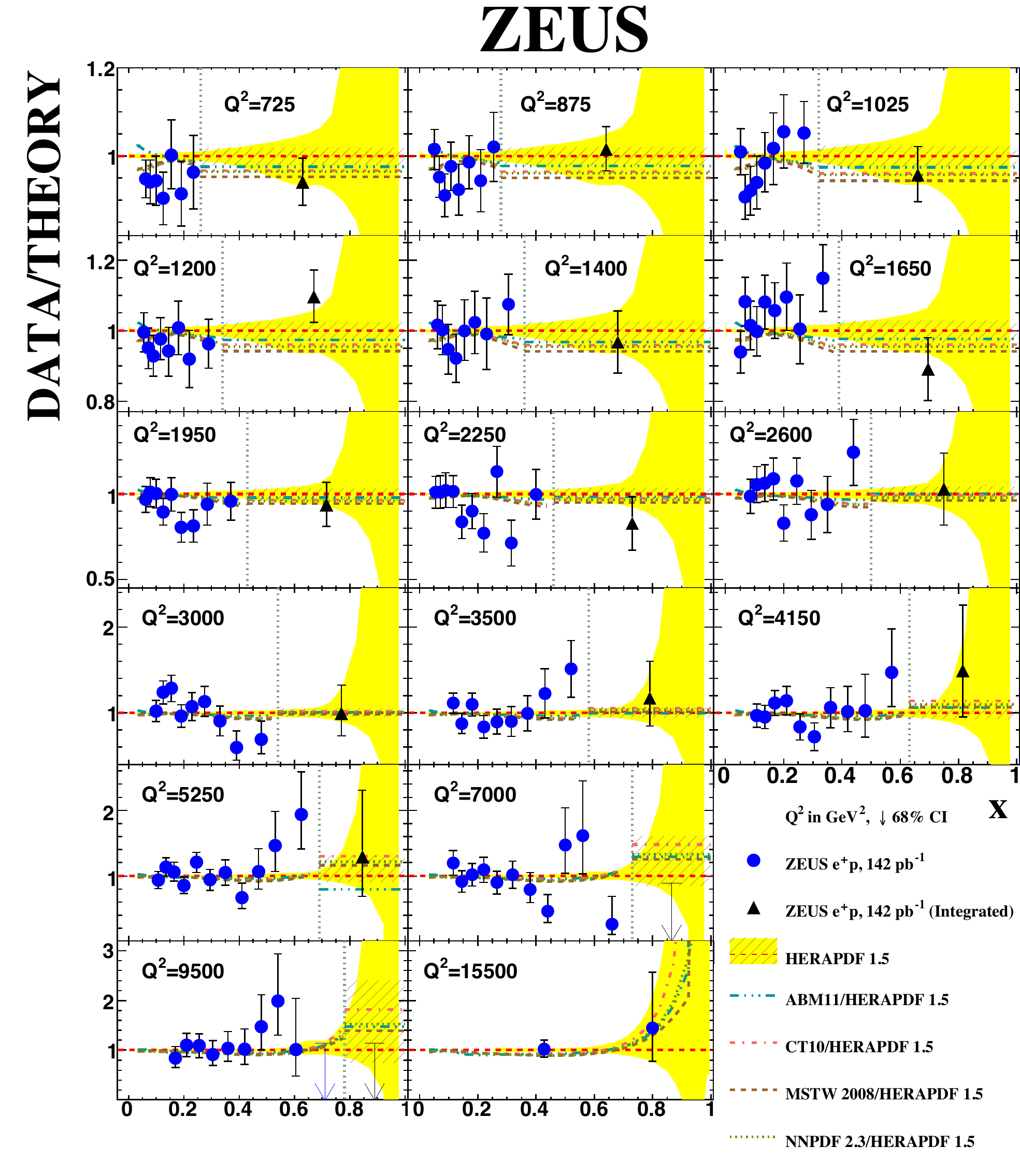}
\caption{
Ratio of the double-differential cross section for NC $e^+p$ 
scattering and  of the double-differential
cross section integrated over $x$ 
to the Standard Model expectation evaluated using the HERAPDF1.5 PDFs as a function of $x$ at different $Q^2$ values as described in the legend.  For HERAPDF1.5, the uncertainty is given as a band. The expectation for the integrated bin is also shown as a hatched box. 
The error bars show the statistical and systematic uncertainties added in quadrature.
The expectations of other commonly used PDF sets normalised to HERAPDF1.5 PDFs are also shown, as listed in the legend. Note that the scale on the $y$ axis changes with $Q^2$. }
\label{fig:ePp}
\end{minipage}
\end{figure}

 For the integrated highest $x$ bins, the respective average cross sections, defined as
\begin{equation}
I(x) = \frac{1}{1-x_{\rm {edge}}}\int_{x_{\rm {edge}}}^{1}\frac{d^2\sigma(x,Q^2)}{dxdQ^2}dx
\;\; ,
\label{eqn-I(x)}
\end{equation}
were obtained and plotted at $x=(x_\mathrm{edge}+1)/2$. The ratio of the measured cross sections to those expected from HERAPDF1.5~\cite{herapdf1.5} are shown in Figs.~\ref{fig:eMp} and~\ref{fig:ePp}.  Note that for bins where no events were observed, the limit is quoted at $68$\% probability,  neglecting the systematic uncertainty. Also shown are the predictions from a number of other PDF sets (ABM11~\cite{abm11}, CT10~\cite{ct10}, MSTW2008~\cite{mstw2008}, NNPDF2.3~\cite{nnpdf2.3}), normalised to the predictions from HERAPDF1.5.  Within the quoted uncertainties, the agreement between measurements and expectations is good.

\section{High-$x$ data and PDF fits}

Simple $\chi^2$ methods are not appropriate for including these data in PDF fits. Many of the bins have only a few or zero events~\cite{zeushighx}. One should use the number of measured events in the fits. Fitters should provide predictions for the number of events expected in the bins in which event counts are reported, and then use Poisson statistics to calculate the probability for the observed number of events given this expectation~\cite{Allen}.

\section{Summary}

The ZEUS collaboration measured double-differential cross sections for $e^\pm p$ NC DIS events at $Q^2 >$ 725 GeV$^2$ up to $x\cong 1$. Fine binning in $x$ and extension of kinematic coverage up to $x\cong 1$ make the data important input to fits constraining the PDFs in the valence-quark domain.

\section*{Acknowledgments}

This activity was partially supported by the Israel Science Foundation.

\end{document}